*Article*

# Adsorption tuning of polarity and magnetism in AgCr$_2$S$_4$ monolayer

Ranran Li,[1] Yu Wang,[1] Ning Ding,[1] Shuai Dong[1]* and Ming An[1]†

[1] School of Physics, Southeast University, Nanjing 211189, China
* Correspondence: sdong@seu.edu.cn
† Correspondence: amorn@seu.edu.cn

**Abstract:** As a recent successfully exfoliated non van der Waals layered material, AgCrS$_2$ has received a lot of attentions. Motivated by its structure related magnetic and ferroelectric behavior, a theoretical study on its exfoliated monolayer AgCr$_2$S$_4$ has been carried out in the present work. Based on density functional theory, the ground state and magnetic order of monolayer AgCr$_2$S$_4$ have been determined. The centrosymmetry emerges upon two-dimensional confinement and thus eliminates the bulk polarity. Moreover, two-dimensional ferromagnetism appears in the CrS$_2$ layer of AgCr$_2$S$_4$ and can persist up to room temperature. The surface adsorption has also been taken into consideration, which shows a nonmonotonic effect on the ionic conductivity through ion displacement of the interlayer Ag, but has little impact on the layered magnetic structure.

## 1. Introduction

Recently, a large number of new two-dimensional (2D) functional materials have been synthesized and reported, including those with intrinsic ferroelectric and long-range spin orders, which have greatly stimulated people's research enthusiasm in 2D ferroic materials [1-12].

Inspired by graphene, most previous researches focus on 2D layered van der Waals (vdW) materials, whose atomic-thin layers can be easily obtained by mechanical exfoliation due to their weak vdW interlayer bonding [13-15]. With the development of 2D research, more feasible approaches emerge. As a supplement to mechanical cleavage, those methods can artificially open a gap between layers of non-vdW materials through selective etching or ionic intercalation. The 2D layer can then be obtained through post-procedure. The typical representatives prepared by chemical etching and intercalation are MXene and AM$_2$X$_4$, respectively [16, 17]. Since then, non-vdW layered materials soon became another emerging branch of 2D materials, especially for those with intrinsic ferroelectricity, long-range spin orders, or even both. For example, NaCrX$_2$ with adjustable conductivity and ACr$_2$S$_4$ (A = Li, Na, K, Rb) nanosheets with multiferroic properties have been reported recently [18, 19].

AgCrS$_2$ is one such layered material with both long-rang magnetic order and ferroelectricity. This compound was synthesized in 1957 [20]. It is composed of an alternative stacking of edge-sharing octahedra CrS$_2$ layer and Ag ion layer along $c$-axis in a trigonal lattice ($R3m$) around room temperature. When cooled down to $T_N$ (about 40 K), the lattice changes to monoclinic ($Cm$), accompanied by the emergence of in-plane double stripes (DS) antiferromagnetic (AFM) order [21, 22]. The ferroelectricity originates from the off-centering displacement of Ag ions. Several experimental works found that this polarization is closely related to the structural and magnetic transition [21, 22]. Recently, the monolayer AgCr$_2$S$_4$, consisting of a single Ag layer sandwiched between two CrS$_2$ layers, was successfully exfoliated from AgCrS$_2$ bulk [17]. Concerns have been aroused [23, 24], mainly focusing on the magnetic and ferroelectric properties of single layer AgCr$_2$S$_4$. However, the specific structure of monolayer and the possible surface adsorption after peeling, and their effects on material properties have not been well explored.

In this work, based on density functional theory (DFT), the magnetic ground state of bulk AgCrS$_2$ has been checked. Our calculation results on bulk are consistent with recent experimental observations, which not only ensure the feasibility of our calculation, but also pave a solid base for the following study on its monolayer. The structural, magnetic, and electronic properties of AgCr$_2$S$_4$ monolayer have been further studied. Unexpectedly, the polar symmetry inherited from parent phase could not be preserved during optimization. Moreover, ferromagnetic (FM) order appears in the in-plane Cr triangular lattice with relatively weak interplane AFM coupling. The situation changes when hydrogen adsorption is taken into consideration. After adsorption, the intralayer FM and interlayer AFM ground state remains unchanged, but the structural symmetry is altered along with its ferroelectricity and ionic transport behavior.

.2. Methods

Our DFT calculations were performed using Vienna *ab initio* Simulation Package (VASP) [25, 26]. The electronic interactions were described by projector-augmented-wave (PAW) pseudo-potentials with semicore states treated as valence states [27]. The exchange and correlation were treated using Perdew-Burke-Ernzerhof (PBE) parametrization of the generalized gradient approximation (GGA) [28]. To properly describe the correlated electrons, the GGA+$U$ method was adopted and the on-site Hubbard $U_{eff}$ was imposed on Cr's 3$d$ orbitals using the Dudarev approach for all calculations [29]. The plane-wave cutoff energy was set to 500 eV. The Monkhorst−Pack $K$-point meshes were chosen as 2×8×4 and 7×7×1 for bulk and monolayer calculations, respectively. Exchange coefficients and magnetic ground states for the monolayer were estimated based on a 2×4×1 supercell with various magnetic orders. The convergent criterion for the energy was set to $10^{-5}$ eV, and that of the Hellman-Feynman forces during structural relaxation was 0.01 eV/Å.

In the study of the monolayer structure, a vacuum layer of 20 Å was added along the $c$-axis direction to avoid the interaction between two neighboring slices. The possible switching paths between different structure phases were evaluated by the nudged elastic band (NEB) method [30]. To estimate the Curie temperature and the temperature evolution of magnetic properties, the Markov-chain Monte Carlo (MC) method with Metropolis algorithm was employed to simulate the magnetic ordering [31]. The MC simulation was performed on a 40×40 lattice with periodic boundary conditions, and larger lattices were also tested to confirm the physical results. The simulations were performed with 20000 equilibration steps and 80000 averaging steps. All MC simulations are gradually cooled down from the initial disordered state at high temperature to the low temperature under investigation.

3. Results

*3.1. AgCrS$_2$ bulk properties*

As mentioned above, the low-temperature bulk crystal belongs to the *Cm* space group without spatial inversion symmetry. Its intrinsic layered feature is illustrated in Figure 1(a). The Cr$^{3+}$ ions located at the center of CrS$_6$ octahedra form a triangular magnetic lattice within each CrS$_2$ layer. Recently, an in-plane collinear magnetic structure has been reported, showing DS pattern (as shown in Figure 1(c)) with AFM coupling in-between [22, 32]. The magnetic ordering and structural transition occur simultaneously, accompanied by the emergence of ferroelectricity, indicating the strong connection between magnetic, ferroelectric, and structural properties [32-35].

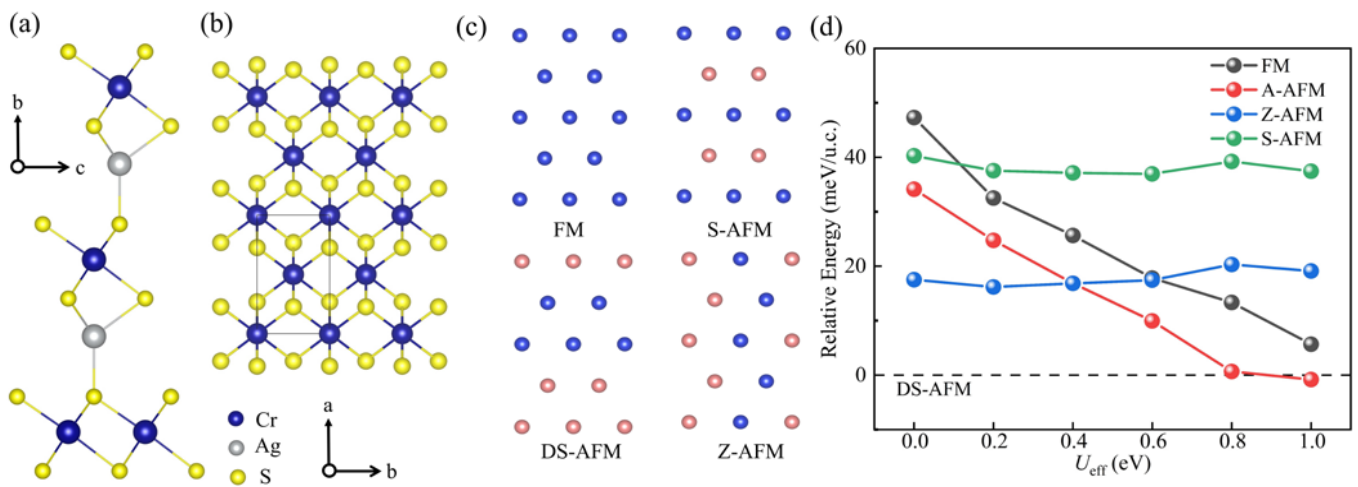

**Figure 1.** (a-b) Side view and top view of bulk AgCrS$_2$. (c) Four in-plane magnetic configurations: ferromagnetic order (FM), stripe AFM (S-AFM), double-stripe order (DS-AFM), and zigzag AFM order (Z-AFM). Blue and red spheres indicate spin-up and spin-down Cr ions, respectively. The other ions are omitted for clarity. (d) The energy evolution of various magnetic orders as a function of $U_{eff}$. The DS state energy is taken as the reference value.

To determine the magnetic ground state, three common collinear configurations on the triangular lattice, as depicted in Figure 1(c), have been taken into account, in addition to the reported DS order. Besides, in order to study the interlayer magnetic coupling, the intralayer FM and interlayer AFM (A-AFM) configuration has also been considered. Our calculation results show that the DS-AFM pattern is indeed the magnetic ground state when $U_{eff}$ is less than 0.8 eV (Figure 1(d)). The total energy of AFM zigzag and stripe configurations are always higher than that of DS-AFM and are almost insensitive to the $U_{eff}$ value. In contrast, the energy of A-AFM and FM decrease with increasing $U_{eff}$, showing a similar trend. The A-AFM order is energetically more favorable than FM, and will even replace DS-AFM as the ground state when $U_{eff}$ is larger than 0.8 eV. Obviously, a specific $U_{eff}$ value (i.e., 0.6 eV) is vital in precisely reproducing $AgCrS_2$ bulk properties. The local magnetic moment increases with the increase of $U_{eff}$, reaching 2.86 $\mu_B$/Cr at 0.6 eV, which is consistent with previous reported value [22]. Moreover, the optimized lattice constants (a = 13.83 Å, b = 3.54 Å, and c = 7.13 Å) are in good agreement with the experimental data [22, 24]. Therefore, it will be adopted in the following calculations by default.

*3.2. $AgCr_2S_4$ monolayer*

Recently, $AgCrS_2$ was successfully exfoliated into 2D nanosheets by Peng *et al*. through ion intercalation [17]. These 2D sheets can be thin down to monolayer, containing one Ag ion layer sandwiched between two $CrS_2$ layers. As can be seen from Figure 2(a, b), the edge-sharing octahedral framework is inherited in $CrS_2$ layers, while the relative displacement of the center Ag ion will give rise to two distinct structural phases (i.e., the asymmetric α phase and the centro-symmetric β phase). The detailed structural information is shown in Table S1 and Figure S1. In the following, we will focus on the structural, magnetic, and electronic properties of $AgCr_2S_4$ monolayer.

To find the ground state of $AgCr_2S_4$ monolayer, the total energies of different magnetic orders mentioned earlier have been calculated based on the above two structural phases. The calculation results have been summarized in Table 1. Obviously, the energy of the β phase is always lower than that of the α phase, presenting a clear tendency to restore the central symmetry of the 2D single layer, different from its parent bulk phase.

Our results also indicate that the magnetic Cr ions in each $CrS_2$ layer tend to couple ferromagnetically and show no sensitivity to the above two structural phases. This structure-insensitive FM behavior, in contrast with its bulk form, can be reasonably interpreted on the basis of the *d* orbital occupation of Cr. In the $AgCrS_2$ bulk, $Cr^{3+}$ ion is in the $3d^3$ configuration. According to the Goodenough-Kanamori-Anderson (GKA) rules [36-38], the half-filled $t_{2g}$ orbitals give rise to AFM direct exchange. While, the *p* orbital intermediated Cr-S-Cr super exchange favors FM coupling. It is the competing exchange interactions that make the bulk magnetic order structurally related. In contrast, from the change of chemical formula before and after exfoliation, the Cr ion in $AgCr_2S_4$ monolayer is in the mixed valence state ($Cr_2^{7+}$). The hole hopping between neighboring Cr's *d* orbitals gives rise to the strong FM tendency and is insensitive to structural details.

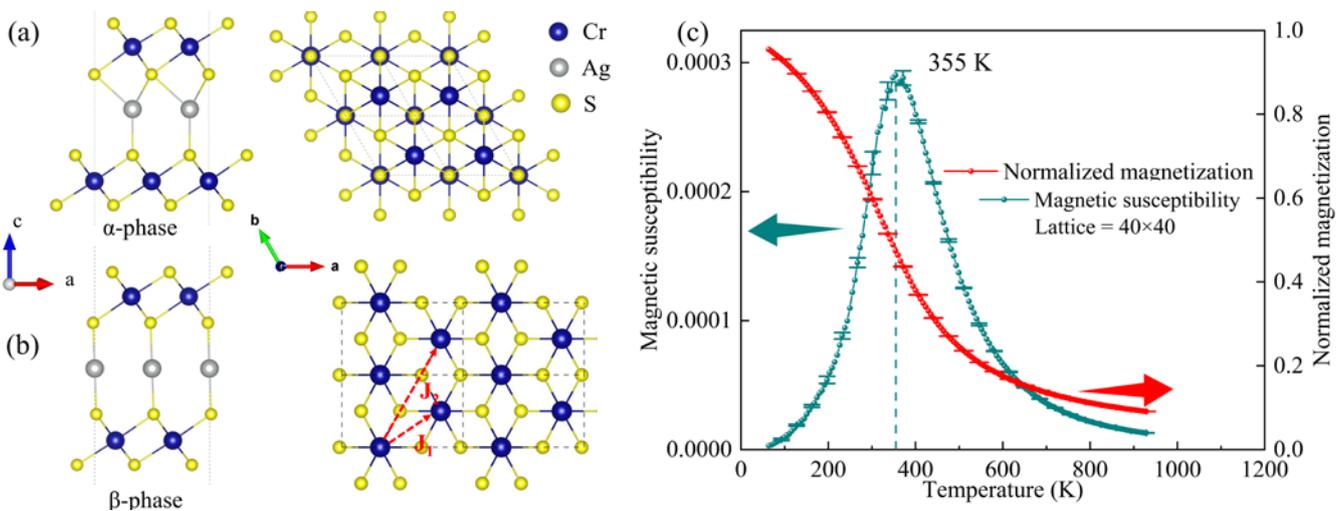

**Figure 2.** (a-b) The top and side views of the $AgCr_2S_4$ asymmetric α phase (*P3m1*) and the centro-symmetric β phase (*C2/m*), respectively. The nearest neighbor interaction $J_1$ and the next-nearest neighbor interaction $J_2$ are denoted by the red dotted arrows in the right panel of (b). (c) The MC simulated magnetic susceptibility and normalized magnetization as a function of temperature for the $AgCr_2S_4$ monolayer.

**Table 1.** The energy differences of four in-plane collinear spin configurations. Interlayer coupling has also been considered, in which the AFM interlayer coupling is denoted by subscript 1 and the FM coupling is represented by subscript 2, respectively. The total energy of the ground state (β phase with A-AFM order) is taken as the reference value, in units of meV/Cr

|   | A-AFM | DS1-AFM | Z1-AFM | S1-AFM | FM | DS2-AFM | Z2-AFM | S2-AFM |
|---|---|---|---|---|---|---|---|---|
| α | 57.25 | 68.40 | 82.62 | 88.25 | 53.65 | 66.44 | 82.82 | 23.39 |
| β | 0 | 14.99 | 11.69 | 24.00 | 2.88 | 13.67 | 13.40 | 22.41 |

To characterize this in-plane triangular magnetic lattice, the classical Heisenberg spin model is adopted, which can be constructed as

$$H = J_1 \sum_{<i,j>} S_i \cdot S_j + J_2 \sum_{[i,k]} S_i \cdot S_k + \sum_i \left[ K_c (S_i^z)^2 + K_b (S_i^y)^2 \right],$$

where $S_i$ is the normalized spin ($|S|=1$) on the Cr site $i$. $J_1$ and $J_2$ correspond to the in-plane exchange constants between the nearest-neighbor (NN) and the next-nearest-neighbor (NNN) interactions, as labeled in Figure 2(b), respectively. $K_{b/c}$ stands for the single-ion magnetocrystalline anisotropy along the $b$-/$c$-axis, respectively. Based on the ground structure (β phase), these exchange coefficients can be extracted by comparing DFT energy with different spin orders. Specifically, in a 2×4×1 supercell, the energy of these magnetic states can be expressed as

$$E_{FM} = E_0 + 3J_1 + 3J_2$$
$$E_{DS} = E_0 + J_1 - J_2 \quad,$$
$$E_Z = E_0 - J_1 + J_2$$

where $E_0$ is the nonmagnetic energy. The derived parameters are summarized in Table 2. According to our estimation, the NN exchange interactions are FM and dominated, as expected in our previous analysis. The NNN exchange constant $J_2$ is relatively weak due to its indirect and long-distance bonding. Based on these exchange parameters, MC simulations were employed to determine the Curie temperature. And the magnetic susceptibility was also calculated. The system reaches equilibrium at a given temperature, the magnetization $M$ and magnetic susceptibility $\chi$ are calculated as [39]:

$$M = \frac{1}{N} \sum_{i=1}^{N} S_i ,$$
$$\chi = \frac{\langle M^2 \rangle - \langle M \rangle^2}{k_B T} ,$$

where $N$ represents the total number of spin sites. Given the exact solution of the spin Hamiltonian, $T_C$ can be estimated from the peak position of the specific magnetic susceptibility $\chi$ (or the maximum slop point of magnetization $M$). MC results are shown in Figure 2(c), indicating that the magnetic transition temperature is above room temperature, much higher than its parent bulk, as expected from its changed and mixed valence of Cr ions. Our MC simulation have also been tested on multi-size lattices to exclude the finite size effect. As shown in Figure S2, the magnetization and susceptibility curve are not sensitive to lattice size, and the Curie temperature shows no obvious scale effect.

Since the AgCr$_2$S$_4$ monolayer contains two CrS$_2$ layers, the interlayer coupling has also been considered. Our calculation shows that the interlayer AFM coupling is energetically more favorable than FM coupling, although the energy difference is quite limited (within 3 meV). After exfoliation, the CrS$_2$-CrS$_2$ interlayer coupling decrease with the increase of the interlayer spacing (from 4.60 Å to 4.76 Å), which is consistent with the intuition. These weakly coupled 2D FM triangular lattices in AgCr$_2$S$_4$ monolayer may provide a new approach for magnetic regulation in spintronic devices.

**Table 2.** In-plane exchange parameters (meV/Cr) estimated from our DFT calculations.

|   | $J_1$ | $J_2$ | $K_b$ | $K_c$ |
|---|---|---|---|---|
| β phase | -14.86 | -5.76 | 1.11 | 0.80 |

The electronic densities of states (DOS) of AgCrS$_2$ bulk and AgCr$_2$S$_4$ monolayer are shown in Figure 3. The parent bulk phase exhibits insulating characteristics with a moderate gap of about 1.5 eV. The states near the Fermi level are mainly originating from Cr's 3$d$ orbitals. In AgCr$_2$S$_4$ monolayer, the stripping induced hole-doping causes a negative shift of the Fermi level, and therefore closes the gap. Meanwhile, Cr's dominant contribution around Fermi level is not affected upon peeling.

In addition, we also verified the mechanical stability of AgCr$_2$S$_4$. The in-plane elastic constants and various mechanical parameters are summarized in Tables S2 and S3, respectively. Our calculation results prove that AgCr$_2$S$_4$ is a soft and malleable material, similar to its three-dimensional counterpart [40]. The schematic diagrams of Young's modulus, shear modulus, and Poisson's ratio are given in Figure S5.

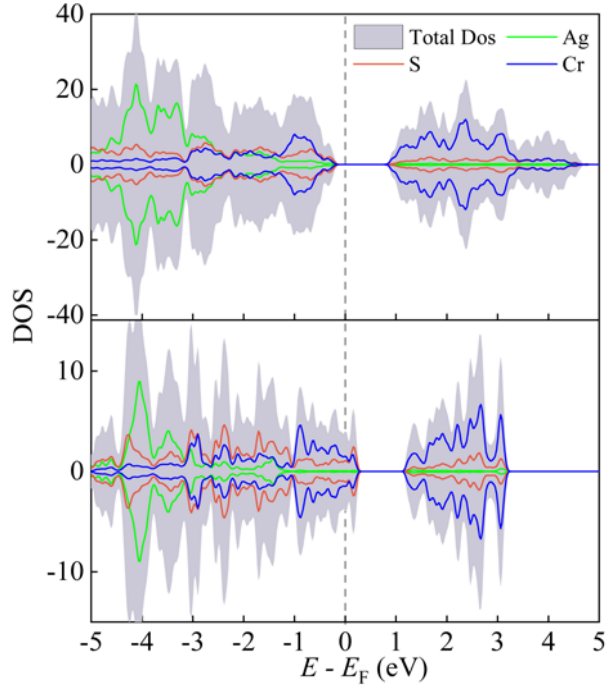

**Figure 3** The density of states (DOS) of bulk AgCrS$_2$ (upper panel) and monolayer AgCr$_2$S$_4$ (lower panel).

### 3.3. *H adsorption effect*

Since the AgCr$_2$S$_4$ monolayer is synthesized through wet chemical exfoliation of bulk AgCrS$_2$. Analogous to MXene, the high surface area to volume ratio and the unsaturated bonds of the outer layer sulfur ions may lead to the adsorption of ions at surface sites, during preparation. Thus, the hydrogen adsorption and its effect on structural and magnetic properties of AgCr$_2$S$_4$ monolayer have been investigated. First, the unilateral passivation is considered. As labeled in Figure 4(a, b), there are ten possible adsorption sites. According to our calculations, the energetically most favorable adsorption site (denoted as C in Figure 4(a)) locates right above the sulfur anion. Detailed adsorption sites and adsorption energy were provided in the supplementary material. After unilateral hydrogen passivation, the central Ag layer shifts away from the adsorption side due to electrostatic repulsion, recovering the original bulk-like AgS$_4$ tetrahedron with neighboring S ions.

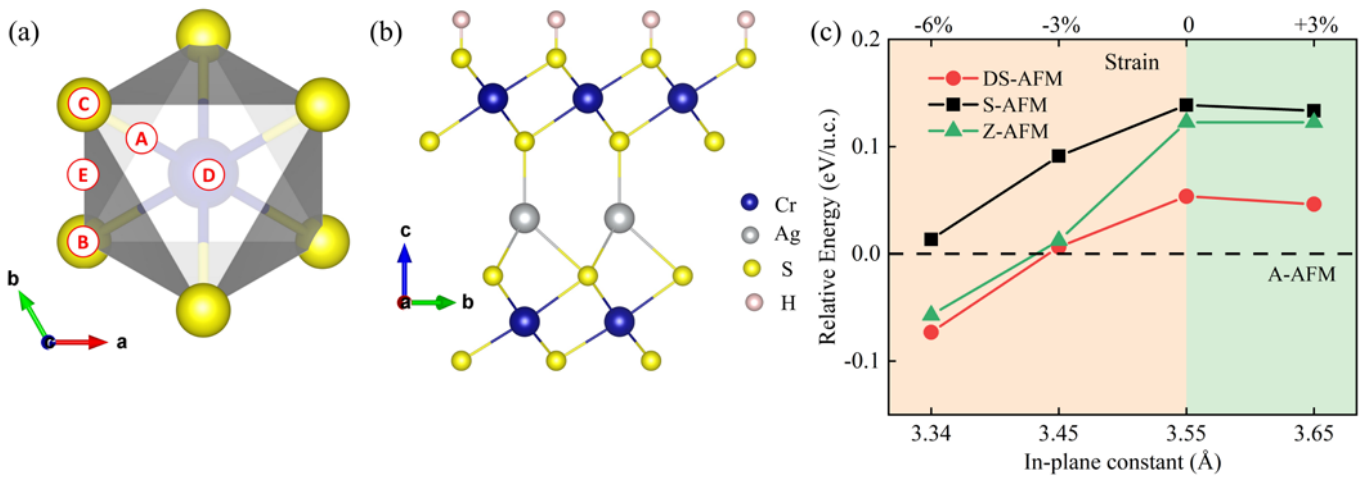

Figure 4. (a) Possible adsorption sites for hydrogen. B, C, D denote the adsorption sites above the lower and upper S ions and Cr ion in the adjacent CrS$_2$ layer, respectively. While, A and E are the adsorption sites located between CD and BC, respectively. (b) Side view of the ground structure of unilaterally passivated AgCr$_2$S$_4$H. (c) AgCr$_2$S$_4$H ground state phase diagram as a function of the lattice constant. Relative energy per unit cell is shown on the left axis. The energy of optimized free-standing structure with A-AFM ground state is taken for reference. Upper axis: the equivalent biaxial strain, defined as $(a - a_0)/a_0$, where $a_0$ and $a$ are the in-plane lattice constants before and after stress application, respectively.

Nonetheless, the magnetic ground state of AgCr$_2$S$_4$H remains A-AFM (e.g. in-plane FM). It is nontrivial since the chemical valence of Cr has been restored to its bulk value (i.e. +3) after the hydrogen adsorption. The in-plane DS-AFM order observed in bulk has not been recovered, which is a little bit unexpected. As discussed previously, the in-plane magnetic pattern in bulk is structurally related, namely the Cr$^{3+}$ ion spacing [22, 35]. We compared the in-plane Cr-Cr spacing after unilateral H passivation with that of the bulk material. The Cr-Cr spacing after passivation is 3.55 Å larger than the bulk value (3.43/3.54 Å, this non-uniform spacing distribution is due to the DS-AFM order). According to GKA rules, the half-filled $t_{2g}$ orbitals of Cr$^{3+}$ give rise to the AFM direct exchange which is sensitive to Cr-Cr spacing. In other words, large spacing weakens this AFM direct exchange, breaks the delicate balance and leads to the FM dominance, which well explains the FM behavior observed in the passivated AgCr$_2$S$_4$H material. Details of lattice constants are marked in Figure S4. This speculation is further proved by in-plane strain modulation. As shown in Figure 4 (c), the in-plane magnetic ground state is fragile and extremely sensitive to biaxial strain. The compressive strains can effectively shorten the in-plane Cr-Cr distance and enhance the direct AFM coupling, thus give rise to the bulk-like DS order. On the contrary, the tensile strains will fasten the in-plane FM order (i.e. A-AFM).

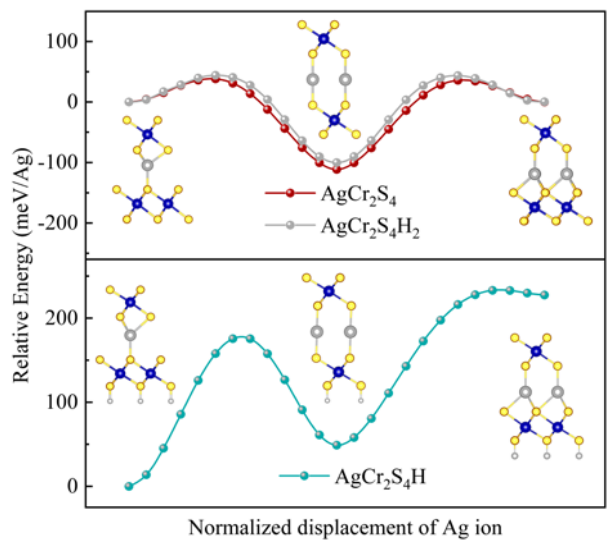

Figure 5. The switching paths of AgCr$_2$S$_4$H$_2$ (grey), AgCr$_2$S$_4$H (blue), and AgCr$_2$S$_4$ (red) as simulated by NEB. Ag, Cr, S, and H ions are represented by gray, dark blue, yellow, and white balls in the illustrations, respectively.

In AgCrS$_2$ bulk, the sandwiched Ag layer has been proved to be crucial to its structure, ferroelectricity, and ionic conductivity [17, 20, 33]. Here, in AgCr$_2$S$_4$ monolayer, Ag's displacement to the central site restores the centrosymmetry and destroys the ferroelectricity. This displacement not only increases the CrS$_2$ interlayer distance, but also weakens the binding between Ag ion and the upper/lower CrS$_2$ layers (Ag-S bonds are halved from 4 to 2), which will certainly lead to the enhancement of ionic mobility as found in experiment [20].

To confirm this scenario, the possible displacement processes of Ag ion in AgCr$_2$S$_4$ as well as its H-passivated cases are simulated by the NEB method. The corresponding energy profiles are shown in Figure 5. Obviously, unilateral adsorption breaks the centro-symmetry and forces Ag to shift away from H, resulting in an asymmetric potential profile, which is consistent with our previous analysis. The simulation results of AgCr$_2$S$_4$ monolayer and the bilaterally adsorbed AgCr$_2$S$_4$H$_2$ are also presented in Figure 5 for comparison. In both AgCr$_2$S$_2$ and AgCr$_2$S$_4$H$_2$, the sandwiched Ag ion tends to locate in the central site with weak bonding between neighboring S ions. The energy barriers of AgCr$_2$S$_2$ and AgCr$_2$S$_4$H$_2$ are 150 meV/Ag and 140 meV/Ag, respectively, much lower than the AgCr$_2$S$_4$H case (190 meV/Ag). For comparison, the energy barrier of AgCrS$_2$ bulk was estimated to be 450 meV/Ag [17].

Based on the above analysis and numerical data, it is reasonable to conclude that the ionic conductivity of AgCrS$_2$ can benefit from dimension reduction and may reach its peak in single-layer AgCr$_2$S$_4$ or its AgCr$_2$S$_4$H$_2$.

## 4. Conclusion

In summary, the structural, magnetic, and electronic properties of bulk AgCrS$_2$ and its single layer AgCr$_2$S$_4$ have been investigated. The in-plane DS pattern of bulk material has been confirmed, but its monolayer exhibits a metallic in-plane FM behavior. The underlying mechanism is attributed to the hole doping induced by exfoliation and the resulting change of Cr's chemical valence. Moreover, the displacement of sandwiched Ag ion towards the high-symmetry center was found, which eliminates the polarity and enhances the ionic conductivity. This structure related superionic behavior is sensitive to surface adsorption. Specifically, it will be inhibited by unilateral adsorption, but will be recovered by bilateral adsorption. The present study may stimulate further experimental and theoretical research on AgCr$_2$S$_4$ and other 2D non-vdW materials.

**Supplementary Materials:** The following supporting information can be downloaded at: https://www.mdpi.com/article/10.3390/ma16083058/s1, Figure S1: In-plane structural parameters of phases α and β; Figure S2: MC simulations of the magnetic susceptibility and normalized magnetization of AgCr$_2$S$_4$ monolayer at different lattice sizes as a function of temperature; Figure S3: Hydrogen adsorption energy and possible adsorption sites of AgCr$_2$S$_4$H; Figure S4: Distance between the Cr-Cr of AgCrS$_2$ and AgCr$_2$S$_4$H; Figure S5: Young's modulus, shear modulus and Poisson's ratio of AgCr$_2$S$_4$; Table S1: The lattice constants of AgCr$_2$S$_4$; Table S2: The elastic constants of AgCr$_2$S$_4$; Table S3: Mechanical parameters of AgCr$_2$S$_4$; Table S4: The energy of AgCr$_2$S$_4$H$_2$ in different magnetic sequences.

**Author Contributions:** Conceptualization, S.D.; methodology, R.R.L; validation, M.A. and S.D.; formal analysis, R.R.L.; resources, S.D. and M.A.; writing—original draft preparation, R.R.L; writing—review and editing, M.A. and S.D.; visualization, R.R.L. and Y.W.; supervision, M.A., S.D. and N.D.; project administration, M.A. and S.D. All authors have read and agreed to the published version of the manuscript.

**Funding:** This research was funded by the National Science Foundation of China (Grant No. 12274069, 12274070).

**Informed Consent Statement:** Not applicable.

**Data Availability Statement:** Data supporting these findings are available from the corresponding authors upon request.

**Acknowledgments:** We thank the Big Data Center of Southeast University for providing the facility support on the numerical calculations.

**Conflicts of Interest:** The authors declare no conflict of interest.

## Appendix A

The following is a list of symbols appeared in the text:

$U_{eff}$: Effective Hubbard parameter. In order to deal with the strong correlation of Cr$^{3+}$, Hubbard empirical parameter $U$ is added using Dudarev's approach, $U_{eff} = U - J$, where $U$ and $J$ are the on-site Coulomb interaction and the strength of the effective on-site exchange interaction, respectively.

$S_i$: The normalized spin vector sitting on lattice site $i$.

$J_1$ and $J_2$: The in-plane nearest neighbor and the next-nearest neighbor magnetic exchange coupling parameters.

$k_B$: The Boltzmann constant.

$M$: The normalized magnetization.

$\chi$: The magnetic susceptibility.